\documentclass[a4paper,11pt]{article}
\pdfoutput=1 

\usepackage{jheppub} 

\usepackage[T1]{fontenc} 
\usepackage{graphics}
\usepackage{graphicx}
\usepackage{dcolumn}
\usepackage{bm}
\usepackage{mathrsfs}
\usepackage{pstricks}
\usepackage{color}
\usepackage{natbib}
\usepackage[normalem]{ulem}
\usepackage{hyperref}
\usepackage{slashed}
\usepackage{amsmath}
\usepackage{epsfig}
\usepackage{amsfonts}
\usepackage{amssymb}
\usepackage{dsfont}
\usepackage{ulem}
\usepackage{wrapfig}
\def\beq{\begin{equation}}
\def\eeq{\end{equation}}
\def\bea{\begin{eqnarray}}
\def\eea{\end{eqnarray}}
\def\nn{\nonumber}
\title{\boldmath Boundary effects on radiative processes of two entangled atoms}

\author[a]{E. Arias,}
\author[b,1]{J. G. Due\~nas,\note{Corresponding author.}}
\author[c]{G. Menezes,}
\author[d]{N. F. Svaiter}
\affiliation[a]{Instituto Polit\'ecnico, Universidade do Estado do Rio de Janeiro, 28625-570 Nova Friburgo, Brazil.}
\affiliation[b]{Universidade Federal de Minas Gerais, Belo Horizonte, BH 31270-901, Brazil.}
\affiliation[c]{Grupo de F\'isica Te\'orica e Matem\'atica F\'isica, Departamento de F\'isica, Universidade Federal Rural do Rio de Janeiro, Serop\'edica, RJ 23897-000, Brazil.}
\affiliation[d]{Centro Brasileiro de Pesquisas F\'{\i}sicas, Rio de Janeiro, RJ 22290-180, Brazil.}

\emailAdd{earias@iprj.uerj.br}
\emailAdd{jgduenas@fisica.ufmg.br}
\emailAdd{gabrielmenezes@ufrrj.br}
\emailAdd{nfuxsvai@cbpf.br}

\abstract{
We analyze radiative processes of a quantum system composed by two identical two-level atoms
interacting with a massless scalar field prepared in the vacuum state in the presence of perfect reflecting flat mirrors.
We consider that the atoms are prepared in a stationary maximally entangled state.
We investigate the spontaneous transitions rates from the entangled states to the collective ground state induced by vacuum fluctuations.
In the empty-space case, the spontaneous decay rates can be enhanced or inhibited depending on the specific entangled state and changes with the distance between the atoms.
Next, we consider the presence of perfect mirrors and impose Dirichlet boundary conditions on such surfaces.
In the presence of a single mirror the transition rate for the symmetric state undergoes a slight reduction, whereas for the antisymmetric state our results indicate a slightly enhancement. Finally, we investigate the effect of multiple reflections by two perfect mirrors on the transition rates.}

\keywords{Decay transition rates, Boundary effects, Quantum entanglement.}
\arxivnumber{1510.00047}

\begin{document}
\maketitle
\flushbottom

\section{Introduction}
\label{intro}
\quad

Superposition of quantum states and quantum entanglement are properties that distinguish quantum mechanics from any classical theory. Entangled states of a system cannot be factorized into product of states of their respective subsystems. This feature clearly exhibits the non-local nature of quantum mechanics. Moreover, entanglement resources have became of interest since it is a key property in quantum information, quantum cryptography and quantum computation~\cite{1,haroche,fi,rep,petrosyan}. Many proposals to generate entangled states in systems of two-level atoms interacting with a bosonic field can be found in~\cite{2,4,5,ved}.

In the semiclassical theory the spontaneous emission of atoms is attributed to the radiation reaction of an oscillating dipole. On the other hand, the quantization of the electromagnetic field leads one to the concept of vacuum fluctuations. In fact, radiation reaction and vacuum fluctuations provide complementary pictures for the interpretation of spontaneous decay of atoms, depending of the particular ordering chosen for commuting atomic and field operators~\cite{aga,milloni}. Here we assume a framework where the spontaneous decay of atoms is only attributed to vacuum fluctuations effects.

By using time-dependent perturbation theory in first-order approximation, it can be shown that the transition rate of an atom interacting with a quantized electromagnetic field in the vacuum state is given by the Fourier transform of the positive frequency Wightman function evaluated on the world line of the atom. In this framework, the probability of transition per unit proper time of a two-level atom from an excited state to the lower-energy state is induced by the vacuum fluctuation of the field on its world line~\cite{birrel,Davies,Unruh,Dewitt,nami}. Although radiative processes are forbidden for an inertial atom prepared in the ground state interacting with the field in the Minkowski vacuum, for a more general trajectory the asymptotic probability of transition can be different from zero. It is well known that an atom moving with constant proper acceleration has a non-null asymptotic probability to undergo a transition to the excited state. This is the Unruh-Davies effect \cite{Davies,Unruh}.

Since the fundamental work of Purcell and Kleppner on the enhancement and inhibition of spontaneous transition rates of atoms inside a resonant cavity~\cite{Purcell,Kleppner}, cavity quantum electrodynamics (CQED) have become an important research field for fundamental investigations and practical applications~\cite{Haroche}. By using the techniques of CQED some approaches have been realized in order to investigate accelerated atoms and the Unruh-Davies effect in cavities~\cite{Scully,Belyanin}. In turn, one can conceive a context in order to study entangled atoms coupled with vacuum fluctuations confined in a cavity. Since the vacuum fluctuations are affected by the presence of the boundaries, one should expect that the atomic transition rates are modified in this scenario~\cite{mesh,ford}.

In this paper, we are interested to analyze how the presence of boundaries affects radiative processes of entangled states. Radiative processes of entangled states have been systematically investigated in the literature. For a careful discussion about radiative process of entangled states see~\cite{fi,rep,petrosyan}. In~\cite{yang} the authors investigate the radiative processes of emission from two entangled atoms coupled with an electromagnetic field in unbounded space. A different scenario was discussed in~\cite{eberly}. These authors study the radiative processes of entangled two-level atoms coupled individually to two spatially separated cavities. The key point of this situation is that each atom indivually interact with vacuum fluctuations inside of each cavity. One can imagine another scenario in which the two entangled atoms interact with a scalar field defined inside only one cavity. It is interesting to ask how the transition rates of entangled atoms are modified by the presence of boundaries~\cite{jose}.

The main purpose of the present work is to analyze quantitatively the effects of boundaries on the transition rates of entangled atoms. We assume two identical two-level atoms coupled with a massless scalar field in Minkowski space-time. The organization of the paper is as follows. In section II we discuss the Hamiltonian describing a system of entangled atoms interacting with the scalar field. We present the spontaneous emission rate in empty space. In section III we evaluate the transition rates of this system in the presence of an infinite reflecting plane. In section IV we generalize our results to the case of two infinite perfect reflecting planes. Conclusions and final remarks are presented in section V. In this paper we use units $ \hbar = c = k_B = 1$.

 \section{Transition rates for entangled atoms in empty space}
\label{model}
\quad

Let us begin considering a single two-level atom coupled with a massless scalar field in a four-dimensional Minkowski space-time~\cite{Dewitt,birrel} .
The atom follows an inertial world-line $x(\tau)$, where $\tau$ is the atom's proper time. The atom-field interaction is described by the usual interaction Lagrangian $g\,m(\tau)\varphi[x(\tau)]$, where $g \ll 1$ is a coupling constant and $m$ is the atom's monopole moment operator. Suppose that the field is initially in the Minkowski vacuum state $|0_M\rangle$ whilst the atom is in the state $|\omega_0\rangle$. By using time-dependent perturbation theory in first-order approximation, one obtains the transition probability amplitude to the final atom-field state $|\omega,\phi_f\rangle$:
\beq
{\cal A}_{|\omega_0,0_M\rangle\rightarrow|\omega,\phi_{f}\rangle}=ig\, \langle\,\omega;\phi_{f}|
\int_{-\infty}^{\infty}\!d\tau\,m(\tau)\varphi[x(\tau)]|0_M;\omega_0\rangle.
\eeq
Within the interaction representation, one has
\beq
m(\tau) = e^{iH_0\tau}m(0)e^{-iH_0\tau},
\label{inter}
\eeq
where $H_0$ is the free Hamiltonian of the single atom. If we consider that $|\omega_0\rangle$  and $|\omega\rangle$ are stationary energy states of the atom, the probability of the atomic transition $|\omega_0\rangle\rightarrow|\omega\rangle$, for any final field configuration, is given by
\beq
P_{|\omega\rangle \rightarrow |\omega_0\rangle}=g^2 \left|\langle \omega|m(0)|\omega_0\rangle\right|^2 F(\omega - \omega_0),
\label{pro}
\eeq
where the response function reads
\beq
F(\omega - \omega_0) = \int_{-\infty}^{\infty}\,d\tau\int_{-\infty}^{\infty}\,d\tau'e^{-i (\omega - \omega_0)(\tau-\tau')} G^{+}[x(\tau),x(\tau')].
\label{resp}
\eeq
In the above equation $G^+[x(\tau),x(\tau')] = \langle 0_M| \varphi(x)\varphi(x')|0_M \rangle$ is the positive frequency Wightman function. The remaining factor in the right-hand side of Eq.~(\ref{pro}) represents the selectivity of the atom which depends on the atomic internal structure.

Let us now investigate the case of two identical two-level atoms interacting with a massless scalar field. 
For simplicity we assume that both atoms remain at rest and we assume that there is not a direct interaction between them. We employ a similar procedure as developed in Ref.~\cite{rep}, namely the Hamiltonian consisting of these two uncoupled identical two-level atoms can be suitably diagonalized and the resulting energies and corresponding eigenstates of the two-atom system are given by~\cite{fi,rep}
\bea
&&E_e = \omega_0\,\,\,\,\,\,\,\,\,\,\,\,|e\rangle = |e_1\rangle|e_2\rangle,
\nn\\
&& E_{ge}= 0\,\,\,\,\,\,\,\,\,\,\,\,\,|ge\rangle = |g_1\rangle|e_2\rangle,
\nn\\
&&E_{eg} = 0\,\,\,\,\,\,\,\,\,\,\,\,\,|eg\rangle = |e_1\rangle|g_2\rangle,
\nn\\
&&E_g = -\omega_0\,\,\,\,\,\,\,|g\rangle = |g_1\rangle|g_2\rangle,
\label{sta}
\eea
where $|g_1\rangle$ and $|g_2\rangle$ are the ground states of the isolated atoms,
and $|e_1\rangle$ and $|e_2\rangle$ are the respective excited states.
In the expressions above $\omega_0$ is the energy gap between the individual atoms states.
The eigenstates of Eq. (\ref{sta}) are known as the product states of two non-interacting atoms. Instead of working with this product-state basis, we can conveniently choose the Bell state basis. In terms of the product states, one has:
\bea
|\Omega^{\pm}\rangle &=& \frac{1}{\sqrt{2}}\left(|e_1\rangle|g_2\rangle \pm |g_1\rangle|e_2\rangle\right)
\nn\\
|\Phi^{\pm}\rangle &=& \frac{1}{\sqrt{2}}\left(|g_1\rangle|g_2\rangle \pm |e_1\rangle|e_2\rangle\right).
\label{bell}
\eea
The Bell states are known as the four maximally entangled two-qubit Bell states, and they form a convenient basis of the two-qubit space. In view of the degeneracy associated with the eigenstates $|ge\rangle$ and $|eg\rangle$, any linear combination of these degenerate eigenstates is also an eigenstate of the atomic Hamiltonian corresponding to the same energy eigenvalue. Therefore, the Bell states $|\Omega^{\pm}\rangle$ are eigenstates of $H_A$. In this work we only consider the entangled states 
$|\Omega^{\pm}\rangle$. Henceforth, we conveniently denote such Bell states as $|\Omega^{+}\rangle = |s\rangle$ and $|\Omega^{-}\rangle = |a\rangle$, with respective energies $E_s = 0 = E_a$. This notation is to better highlight the (anti)symmetric nature of the state ($|a\rangle$) $|s\rangle$.

The interaction Lagrangian between each atom and the field is given by
$g\,m_1(\tau)\varphi[x_1(\tau)]$ and $g\,m_2(\tau)\varphi[x_2(\tau)]$.
These terms depend implicitly on each of the atomic world-lines, $x_1(\tau)$ and $x_2(\tau)$.
Here the operators $m_1$ and $m_2$ are the monopole moments of each atom expressed in the extended Hilbert space of the two atoms, i.e., $m_1=m\otimes\mathds{1}_{2}$ and $m_2=\mathds{1}_{1}\otimes m$, $m$ being the monopole moment operator of the isolated atoms. In order to analyze the transition rates of system, let us assume that the field is in the Minkowski vacuum state $|0_M\rangle$ and the two-atoms system is in a state $|\omega'\rangle$. Then the transition probability to the collective state $|\omega\rangle$ for the atoms reads
\bea
P_{|\omega'\rangle \to |\omega\rangle} &=& g^2 \bigg[|m_{\omega \omega'}^{(1)}|^2 F_{11}(\Delta\omega)
+ |m_{\omega \omega'}^{(2)}|^2 F_{22}(\Delta\omega)\nn\\
&+&m_{\omega \omega'}^{(1)}\,m_{\omega \omega'}^{(2)*}F_{21}(\Delta\omega)
 +m_{\omega \omega'}^{(2)}\, m_{\omega \omega'}^{(1)*}F_{12}(\Delta\omega)\bigg],
\label{pro1}
\eea
where we have defined $\Delta\omega=\omega - \omega'$ and the matrix elements are given by
\bea
m_{\omega \omega'}^{(1)} &=& \langle\omega|m\otimes\mathds{1}_{2}|\omega'\rangle
\nn\\
m_{\omega \omega'}^{(2)} &=& \langle\omega|\mathds{1}_1\otimes m|\omega'\rangle.
\label{mel}
\eea
In Eq. (\ref{pro1}) we have considered that
the states $|\omega\rangle$ and $|\omega'\rangle$ belong to the collective set $\{|g\rangle, |a\rangle, |s\rangle,  |e\rangle\}$,
discussed above. Respectively $\omega$ and $\omega'$ can be any of the atomic energies $\{E_g, E_a, E_s, E_e\}$.
The corresponding response functions are given by
\beq
F_{ij}(\Delta\omega) = \int_{-\infty}^{\infty}\,d\tau\int_{-\infty}^{\infty}\,d\tau' e^{-i \Delta\omega(\tau-\tau')} G^{+}[x_i(\tau),x_j(\tau')],
\eeq
where $i,j=\{1,2\}$ and $G^{+}[x_i(\tau),x_j(\tau')] = \langle 0_M| \varphi(x_i(\tau))\varphi(x_j(\tau'))|0_M \rangle$.
We see from Eq. (\ref{pro1}) that the transition probability of the two-atoms system presents contributions from the isolated atoms, $F_{11}$ and $F_{22}$, and also contributions due to cross-correlations between the atoms mediated by the field, $F_{12}$ and $F_{21}$. This interference is a consequence of the interaction of each atom with the field. The information of the entangled state is encoded in the matrix elements $m_{\omega \omega'}^{(i)}$, $i = 1, 2$. Let us discuss the transition probability per unit proper time. For the general transition of the two-atom system from $|\omega'\rangle$ to $ |\omega\rangle$, we obtain the following transition rate
\bea
{\cal R}_{|\omega'\rangle \to |\omega\rangle} &=& g^2 \bigg [|m_{\omega \omega'}^{(1)}|^2
{\cal F}_{11}(\Delta\omega)
+ |m_{\omega \omega'}^{(2)}|^2
{\cal F}_{22}(\Delta\omega) \nn\\
&+&m_{\omega \omega'}^{(1)}\,m_{\omega \omega'}^{(2)*}
{\cal F}_{21}(\Delta\omega)
+ m_{\omega \omega'}^{(2)}\,m_{\omega \omega'}^{(1)*}
{\cal F}_{12}(\Delta\omega)
\bigg],
\label{pro2}
\eea
where the response function per unit time is defined as
\beq
{\cal F}_{ij}(\Delta\omega) =
\int_{-\infty}^{\infty}\,d(\Delta\tau) e^{-i \Delta\omega\Delta\tau} G^{+}[x_i(\tau),x_j(\tau')],
\label{Fij}
\eeq
where $\Delta\tau=\tau-\tau'$ and $i,j=\{1, 2\}$.
\begin{figure}
\centering\includegraphics[width=0.55\linewidth]{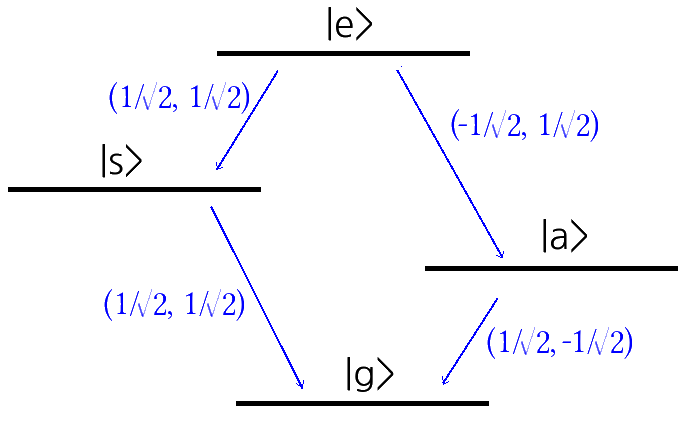}
        \caption{Possible transitions for the two-atom system caused by field vacuum fluctuations. In each transition is indicated the
        the matrix elements of the isolated monopole operators of the atoms $(m^{(1)}_{\omega'\omega},m^{(2)}_{\omega'\omega})$.
        For example, from above one sees that the transition $|e\rangle\rightarrow|a\rangle$ has
        $m^{(1)}_{ae}=-1/\sqrt{2}$ and $m^{(2)}_{ae}=1/\sqrt{2}$.
        The direct transition $|e\rangle\rightarrow|g\rangle$ has null monopole matrix element,
        $m^{(1)}_{ge}=m^{(2)}_{ge}=0$, hence this transition is forbidden and is not represented in the diagram above.}
\label{matrixelements}
\end{figure}
Some remarks about the matrix elements given by Eq.~(\ref{mel}) are in order. Since the monopole matrix of the $i$-th atom is defined as $m_{i}= |e_{i}\rangle\langle g_{i}| + |g_{i}\rangle\langle e_{i}|$, one can calculate the monopole matrix elements for specific transitions
$|\omega' \rangle \to |\omega \rangle$.
The matrix elements that correspond to the transition from the symmetric entangled state to the ground state
are $m_{gs}^{(1)}=m_{gs}^{(2)}=1/\sqrt{2}$.
For the transition from the antisymmetric entangled state to the ground state
one gets $m_{ga}^{(1)}=-m_{ga}^{(2)}=1/\sqrt{2}$.
The transition $| e \rangle \to | g \rangle$ is forbidden  due to selection rules, since
$m_{ge}^{(1)}=m_{ge}^{(2)}=0$.
All the permitted transitions are depicted in Fig. (\ref{matrixelements}).

The positive frequency Wightman function in Minkowski space-time for a massless scalar field is given by \cite{birrel}
\beq
G^{+}[x,x'] = -\frac{1}{4\pi^2}\frac{1}{\left[(t - t' - i\epsilon)^2 - |{\bf x} - {\bf x}'|^2\right]},
\label{wigh}
\eeq
where the space-time points are $x = (t,{\bf x})$ and we have introduced an infinitesimal positive parameter $\epsilon$ to specify the singularities of the function. Since we consider that the atoms remain at rest, they will follow inertial world lines $x_{i}(\tau)=(\tau,{\bf x}_i)$. In this way the space-time interval between the atoms at different proper times is $\Delta x = x_1(\tau) - x_2(\tau')=(\Delta\tau , {\bf d}_{-})$, where we have defined the relative position vector between the atoms as ${\bf d}_{-}= {\bf x}_{1} - {\bf x}_{2}$. Hence the Wightman functions in Eq. (\ref{Fij}) are given respectively by
\bea
&& G^{+}[x_i(\tau),x_j(\tau')] = -\frac{1}{4\pi^2}\frac{1}{(\Delta\tau - i\epsilon)^2},\,\,\, (i=j),
\nn\\
&& G^{+}[x_i(\tau),x_j(\tau')] = -\frac{1}{4\pi^2}\frac{1}{\left[(\Delta\tau - i\epsilon)^2 - |d_{-}|^2\right]},\,\,\, (i\ne j).
\eea
The integrals in the right-hand side of Eq.~(\ref{Fij}) and as well as some others that arise along this work can be performed using the residue's theorem. These are of the form
\beq
-\frac{1}{4\pi^2}\int_{-\infty}^{\infty}\,dy \frac{e^{-i by}}{(y - i\epsilon)^2-z^2} = -\frac{\theta[-b]}{2\pi}\frac{\sin(bz)}{z},
\label{intresidues}
\eeq
where $b$ and $z$ are constants and $\theta$ is the Heaviside step function. Hence, the total transition rate of the two-atom system in empty space is given by
\bea
{\cal R}_{|\omega'\rangle \to |\omega\rangle}&=& \frac{g^2}{2\pi}\,\theta(-\Delta\omega)|\Delta\omega|\biggl[|m_{\omega \omega'}^{(1)}|^2+
|m_{\omega \omega'}^{(2)}|^2 \nn\\
&+&\big(m_{\omega \omega'}^{(1)}\,m_{\omega \omega'}^{(2)*} + m_{\omega \omega'}^{(2)}\,m_{\omega \omega'}^{(1)*}\big)
\frac{\sin\left(\Delta\omega|d_{-}|\right)}{\Delta\omega|d_{-}|}\biggr].
\label{gamma-total}
\eea
Observe that the first two terms above are the expected contributions from the individual atomic transitions.
However the energy gap is that of an entangled state.
The other two terms in Eq. (\ref{gamma-total}) exhibit the existence of cross-correlations of the field evaluated at the different world-lines of the atoms, $x_1$ and $x_2$.
From Eq. (\ref{gamma-total}) we see that the spontaneous transition rate can be enhanced or inhibited depending on the matrix elements of each transition and the separation between the atoms. The cross-correlations generate an interference pattern in the transition rate which has a similar behavior for both possible transitions. It depends on the distance between the atoms $d_-$, and is characterized by the wavelength $\lambda=2\pi/|\Delta\omega|$ associated with the transition energy gap.

\begin{figure}
\centering\includegraphics[width=0.6\linewidth]{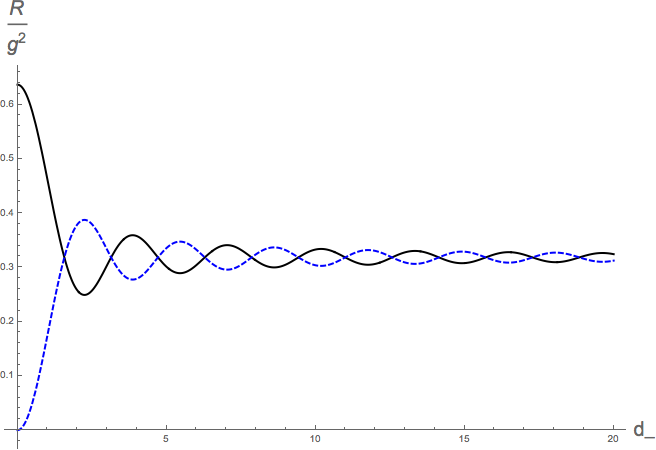}
        \caption{Spontaneous transition rate for two-level entangled atoms at rest separated a distance $|d_{-}|=|d_1 - d_2|$.
        The two-atoms system decay from  the symmetric (continuous line) and from the antisymmetric (dashed line) state
         to the ground collective state.
         We take the energies $\omega_0=2.0$, measured in units of $2\pi\lambda_{gw}^{-1}$. For simplicity we orient the z-axis along the line joining the two point-like atoms.}
\label{fig:2LAent}
\end{figure}

\textit{Symmetric state transition}: In order to see how the transition rates  of the two-atoms system can be enhanced or inhibited,
let us consider specifically the transition $| s \rangle \to | g \rangle$.
By using the corresponding matrix elements of this transition $m_{gs}^{(1)}=m_{gs}^{(2)}=1/\sqrt{2}$,
we get the transition rate
\beq
{\cal R}_{|s\rangle \to |g\rangle} =  \frac{g^2}{2\pi}\,|E_{gs}|
\left[1
+
\frac{\sin\left(|E_{gs}||d_{-}|\right)}{|E_{gs}||d_{-}|}\right],
\label{gamma-sg}
\eeq
where $|E_{gs}|= |E_g - E_s| = \omega_0$.
Because of the positive matrix elements, 
the probability transition rate is increased in comparison with the case where the entangled atoms are far enough separated.
The modes of the field which are resonant with this transition are those for which $\omega  \sim |E_{gs}|$.
For this case whenever $E_{sg}d_{-}=(2n+1/2)\pi$, $n$ being a positive integer, one has a constructive interference. It means that if the distance between the atoms is $|d_{-}|=(n+1/4)\lambda_{gs}$, where $\lambda_{gs}=2\pi/|E_{gs}|$, the transition rate gets increased. On the other hand, destructive interference happens for $E_{sg}d_{-}=(2n+3/2)\pi$ which implies a lower transition probability rate for relative distances $|d_{-}| = (n+3/4)\lambda_{gs}$. For $|d_{-}|=n\lambda_{gs}$ these cross-correlations terms vanish. In addition, for small distances between both atoms $|d_{-}|\ll \lambda_{gs}$, there is an increase of the transition rate
by a factor of two in comparison with the case in which the distance between the atoms is very large  $|d_-|\gg\lambda_{gs}$.
It means that the quantum correlations between the atoms mediated by the field generates a constructive interference when
the atoms are near enough  each other and these interference terms vanish for large spatial separations between entangled atoms. It is clear that there is a natural lower bound for this separation given by the sizes of the atoms.
A picture of the behavior described above is illustrated in Fig.~(\ref{fig:2LAent}). 

\textit{Antisymmetric state transition}: The spontaneous transition rate for the two-atom system decay $| a \rangle \to | g \rangle$ can be obtained in a similar way. In this case, recalling that the matrix elements are $m_{ga}^{(1)}=-m_{ga}^{(2)}=1/\sqrt{2}$,
we obtain:
\beq
{\cal R}_{|a\rangle \to |g\rangle} = \frac{g^2}{2\pi}\,|E_{ga}|
\left[1
-\frac{\sin\left(|E_{ga}||d_{-}|\right)}{|E_{ga}||d_{-}|}\right],
\label{gamma-ag}
\eeq
with the energy gap of the transition being $|E_{ga}|= |E_g - E_a| = \omega_0$. Associated with this transition we define the wavelength  $\lambda_{ga}=2\pi/|E_{ga}|$. For the antisymmetric state we have an opposite behavior as compared with the symmetric case. For distances between the atoms such that $|d_{-}|=(n+1/4)\lambda_{ga}$ we get a lower transition rate and for distances $|d_{-}|=(n+3/4)\lambda_{ga}$ the interference operates in order to enhance the transition rate, see Fig.~(\ref{fig:2LAent}). Also, unlike the symmetric state, for small distances $|d_{-}|\ll \lambda_{ga}$, we now have a complete inhibition of the spontaneous transition rate due to destructive interference of quantum correlations between the atoms mediated by the field.  On the other hand, note that for large separations between the atoms the quantum interference effects for both transitions produce vanishing contributions. It implies that the influence of the quantum interference is stronger for short distances between the atoms. For large separations the fact that the system is in an entangled state remains only coded in the energy gap of the transition. This scenario was also considered in Ref.~\cite{LinHu2009}. In such a reference the authors showed that for two detectors at rest interacting with a massless scalar field as their environment, the entanglement dynamics depends on the spatial separation between detectors and vanishes for large distances.

We remark that the Eq.~(\ref{gamma-ag}) is very similar to the results obtained in~\cite{ford}, where the distance between the atoms is replaced by twice the distance between the atom and a mirror boundary and the gap energy is that for a single atom. In the latter situation it also was regarded that the atom is coupled with the vacuum fluctuations of the field rather than interacting directly with its mirror image. We can picture the decay of the antisymmetric state as the decay of a single atom in the presence of an infinite plane interacting with scalar field satisfying Dirichlet boundary conditions, whereas the decay of the symmetric state would be similar to the decay of an atom  in the presence of an infinite plane interacting with scalar field satisfying Neumann boundary conditions. The difference of considering the influence of either a real object or a mirror image on the entanglement dynamics between an atom and a quantum scalar field in the presence of a mirror was analized in~\cite{ZhouHu2012}.

\section{Transition rates for entangled atoms in the presence of a mirror}
\label{plane}
\quad

In this Section we investigate the transition rates of the two atoms described before in the presence of perfectly reflecting mirrors. Let us assume the case of an infinite plane in unbounded four-dimensional Minkowski space. We impose Dirichlet boundary conditions on the field at the plane's surface $x_3 = 0$, given by
\beq
\varphi (x_3 = 0) = 0.
\label{diri}
\eeq
The positive frequency Wightman function is given by
\bea
G^{+}[x,x'] &=& -\frac{1}{4\pi^2}\Bigg[\frac{1}{(\Delta t - i\epsilon)^2 - \Delta {\bf x}_{\perp}^2 - (x_3 - x_3')^2}
\nn\\
&-&\,\frac{1}{(\Delta t - i\epsilon)^2 - \Delta {\bf x}_{\perp}^2 - (x_3 + x_3')^2}\Bigg],
\label{wighp}
\eea
with $\Delta {\bf x}_{\perp}^2 = (x_1 - x_1')^2+(x_2 - x_2')^2$.
%
As previously, we assume that the atoms are at rest at a distance $d_1$ and $d_2$ from the plane. Respectively, their world-lines are $x_{i}^{\mu}(\tau) = (\tau, 0, 0, d_i)$, for $i=\{1,2\}$. Correspondingly, the Wightman functions in Eq.~(\ref{Fij}) are given by
\bea
&& G^{+}[x_i(\tau),x_j(\tau')] = -\frac{1}{4\pi^2}\Biggl[\frac{1}{(\Delta\tau - i\epsilon)^2}
-\frac{1}{(\Delta\tau - i\epsilon)^2 - (2 d_i)^2}\Biggr],\,\, (i= j),
\\
&& G^{+}[x_i(\tau),x_j(\tau')] = -\frac{1}{4\pi^2}\Biggl[\frac{1}{(\Delta\tau - i\epsilon)^2 - (d_{-})^2}
-\frac{1}{(\Delta\tau - i\epsilon)^2 - (d_{+})^2}\Biggr],\,\, (i\neq j),
\eea
where $d_{+}=d_1+d_2$. Now we insert these results into the right-hand side of Eq. (\ref{pro2}) and perform the integral as indicated in Eq. (\ref{intresidues}). The total transition probability per unit proper time is given by
\bea
{\cal R}_{|\omega'\rangle \to |\omega\rangle} = g^2\,\frac{\theta(-\Delta\omega)|\Delta\omega|}{2\pi} &\Biggl\{&|m_{\omega \omega'}^{(1)}|^2
\left[ 1 - \frac{\sin\left(2d_1 \Delta\omega\right)}{2d_1 \Delta\omega}\right]
+ |m_{\omega \omega'}^{(2)}|^2
\left[ 1 - \frac{\sin\left(2d_2 \Delta\omega\right)}{2d_2 \Delta\omega}\right]\nn\\
&&\!\!\!\!\!\!\!\!\!\!\!\!\!\!\!\!\!\!\!  + \left[m_{\omega \omega'}^{(1)}m_{\omega \omega'}^{(2)*} +m_{\omega \omega'}^{(2)}m_{\omega \omega'}^{(1)*}\right]
\left[\frac{\sin\left(\Delta\omega d_{-}\right)}{\Delta\omega d_{-}} - \frac{\sin\left(\Delta\omega d_{+}\right)}{\Delta\omega d_{+}}\right]\Biggr\}.
\label{gamma-plane}
\eea
 \begin{figure}
\centering\includegraphics[width=0.7\linewidth]{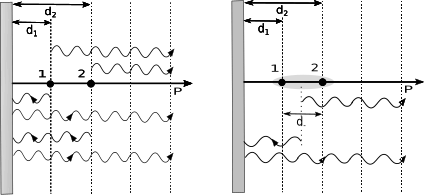}
     \caption{Schematic representation of the quantum interference for two identical two level atoms interacting with a massless scalar field in the presence of one mirror.  For two isolated atoms to each other (\textit{left}), the field detected in the point $p$ far away at right of a mirror is the sum of the emmited radiation by each individually atom and the corresponding radiation due to the reflected waves on the mirror [first line of Eq (3.5)]. If we consider that both atoms form a system (\textit{right}), we must add two contributions to the last case due to the cross-correlations between the atoms mediated by the field. There is one contribution depending on the distance between the atoms $d_{-}$, and other depending on the sum of the distances of atoms to the mirror $d_{+}$ [second line of Eq. (3.5)].} 	
    \label{fig:Diagram}
\end{figure}
A schematic representation of the physical situation is illustrated in Fig.~(\ref{fig:Diagram}). We can interpret this result in a similar way as was done in the previous section. The terms in the first line of Eq.~(\ref{gamma-plane}) correspond to the decay rates of a pair of two-level atoms isolated from each other in the presence of a mirror. These interference terms, as in \cite{ford}, are given by the reflected field on the mirror and depend on the distance of each atom to the mirror.  The next terms in the second line of Eq.~(\ref{gamma-plane}) are the cross-correlation between the atoms due to the entangled state as was shown in the Eq.~(\ref{gamma-total}). The first term depends on the distance between the atoms $d_{-}$, whereas the second term in Eq.~(\ref{gamma-plane}) describes the cross-correlation between the atoms depending on the distance $d_{+}$. It is the distance that a reflected wave on the mirror needs to travel from one atom to reach the other. In a naive way, this is the distance from one atom to the mirror image of the other. However these are two different physical phenomena as has been remarked in ~\cite{ZhouHu2012}. If we compare the Eq.~(\ref{gamma-total}) with the Eq.~(\ref{gamma-plane}), we see that the presence of a mirror generates interference terms in the spontaneous transition rate. For a detailed analysis, now let us consider the two specific transitions from the maximally entangled states of the system to its collective ground state.

\textit{Symmetric state transition}: For the decay of the symmetric state to the collective ground state the transition probability per unit proper time in the presence of an infinite mirror reads
\bea
{\cal R}_{|s\rangle \to |g\rangle} = \frac{g^2|E_{gs}|}{2\pi}\Biggl[
 1 &-& \frac{\sin\left(2d_1|E_{gs}|\right)}{4d_1 |E_{gs}|}
   - \frac{\sin\left(2d_2|E_{gs}|\right)}{4d_2 |E_{gs}|}\nn\\
&+&\frac{\sin\left(d_{-}|E_{gs}|\right)}{d_{-}|E_{gs}|}
- \frac{\sin\left(d_{+}|E_{gs}|\right)}{d_{+}|E_{gs}|}\Biggr].
\label{gamma-plane-sg}
\eea
The presence of the mirror also modifies slightly the sinusoidal behavior of transition rate. Interference contributes to lower the transition rate if the atoms are located at $d_1=d_2=(n+1/4)\lambda_{gs}/2$, $n$ a positive integer; in turn interference effects enhance the transition rate if $d_1=d_2=(n+3/4)\lambda_{gs}/2$.
\begin{figure}
\includegraphics[width=0.53\linewidth]{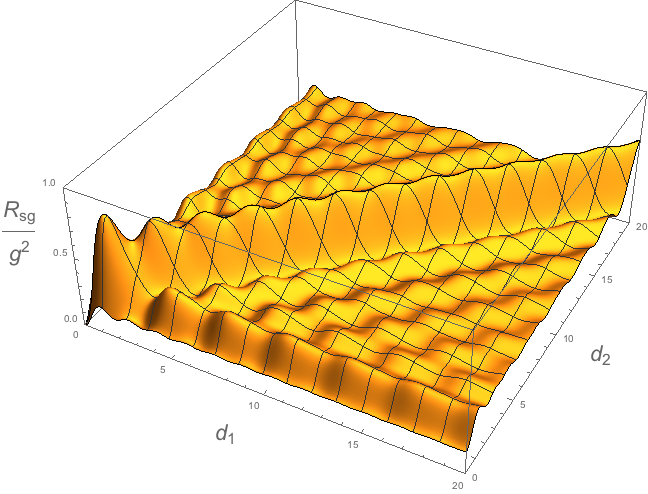}
\includegraphics[width=0.53\linewidth]{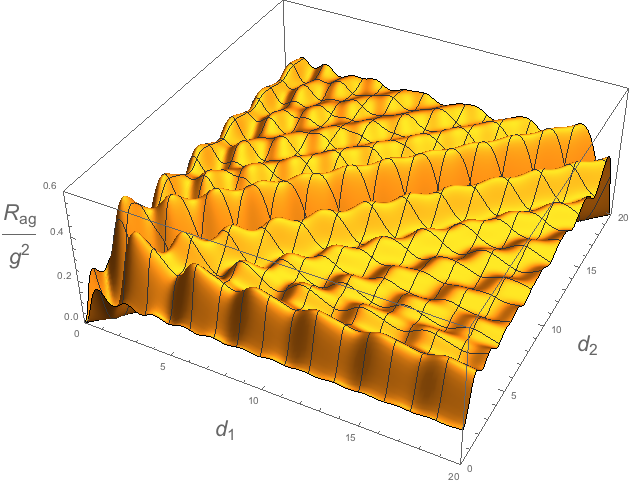}
     \caption{Decay per unit time as function of distances $d_1$ and $d_2$ of the two atoms to the plate. The transition rates are given in units of $g^{-2}$ and the energies and distances are given in terms of the natural units associated to each transition. This means that the distances for ${\cal R}_{gs}$ are in units of $\lambda_{gs}$ whilst the distances for ${\cal R}_{ga}$ are in units of $\lambda_{ga}$.} 	
    \label{fig:DecRatePlate}
\end{figure}

\textit{Antisymmetric state transition}: For the decay of the antisymmetric state to the collective ground state the transition probability per unit proper time in the presence of an infinite mirror one has
\bea
{\cal R}_{|a\rangle \to |g\rangle} = \frac{g^2|E_{ga}|}{2\pi}\Biggl[
1 &-& \frac{\sin\left(2d_1|E_{ga}|\right)}{4d_1 |E_{ga}|}
- \frac{\sin\left(2d_2 |E_{ga}|\right)}{4d_2|E_{ga}|}\nn\\
&-& \frac{\sin\left(d_{-}|E_{ga}|\right)}{d_{-}|E_{ga}|}
+ \frac{\sin\left(d_{+}|E_{ga}|\right)}{d_{+}|E_{ga}|}\Biggr].
\label{gamma-plane-ag}
\eea
We obtain an opposite behavior due to quantum correlations. The contribution depending on the distance between atoms tends to decrease the transition rate. However, the reflection of the field in the mirror tends to enhance the spontaneous transition rate and has the same spatial dependency as in the symmetric case, see Eq. (\ref{gamma-plane-sg}).
For distances such that $d_1=d_2$, the spontaneous transition rate is completely inhibited. In general, the interference pattern is qualitatively similar to the symmetric transition. For instance, if the atoms are located at distances such that $d_1=(n+1/4)\lambda_{ag}/2$ and $d_2$ such that their sum a constant $d_1+d_2=l$, where $l$ is arbitrary, and $d_2 \neq d_1$, the transition rate will be reduced. But if they are located at distances such that $d_1=(n+3/4)\lambda_{ag}/2$ and $d_1+d_2=l$ with $l$ the same constant as before and $d_2 \neq d_1$, the transition rate gets increased.

The transition rates of both symmetric and antisymmetric states to the collective ground state are illustrated in Fig.~(\ref{fig:DecRatePlate}). If we compare the results of Fig.~(\ref{fig:2LAent}) and Fig.~(\ref{fig:DecRatePlate}) we see that the presence of the mirror induces  in the transition rate a slight increasing for the antisymmetric state and a slight decreasing for the symmetric state. The asymptotic behavior of the transition rate remains qualitatively the same as the case without mirror for large distances between the atoms. 

\newpage
\section{Transition rates for entangled atoms in the presence of two mirrors}
\label{planes}
\quad

In this section we extend our studies of radiative processes of two entangled atoms by considering boundaries
that fully confine the quantum field in one spatial direction. We consider the presence of two parallel reflecting planes located along the $x_3$-axis and adopt Dirichlet boundary conditions for the scalar field at the mirror's surfaces
\beq
\varphi (x_3 = 0) = \varphi (x_3 = L) = 0,
\label{diri2}
\eeq
where $L$ is the distance between the two mirrors. The positive frequency Wightman function is given by the series
\bea
G^{+}[x,x'] = -\frac{1}{4\pi^2}\sum_{k = -\infty}^{\infty}&\Biggl[&\frac{1}{(\Delta t - i\epsilon)^2 - \Delta {\bf x}^2_{\perp} - (x_3 - x_3' - kL)^2}
\nn\\
&-&\,
\frac{1}{(\Delta t - i\epsilon)^2 - \Delta {\bf x}^2_{\perp} - (x_3 + x_3' - kL)^2}\Biggr],
\label{wighp2}
\eea
which vanishes at points on the plates at $x_3$ or $x_3' = 0$ and $x_3$ or $x_3' = L$, as required.
In the above expression $\Delta {\bf x}_{\perp}={\bf x}_{\perp}-{\bf x}'_{\perp}$ is the distance between the points perpendicular to the $x_3$-axis. As above, let us consider the two identical atoms at rest on the $x_3$-axis at distances $d_1$ and $d_2$ from the $x_3= 0$ plane. We only consider configurations in which the atoms are placed on the line perpendicular to the mirrors. In this case the world lines are again given by $x_{i}^{\mu}(\tau) = (\tau, 0, 0, d_i)$, for $i=1,2$, respectively. In order to calculate the transition rates of the two-atom system given by Eq. (\ref{pro2}), we need to evaluate the Eq. (\ref{wighp2}), at the world lines of the atoms. Hence  we obtain that
\bea
G^{+}[x_i(\tau),x_j(\tau')]\bigr|_{i=j} &=& -\frac{1}{4\pi^2}\sum_{k = -\infty}^{\infty}\Biggl[\frac{1}{(\Delta\tau - i\epsilon)^2 - (kL)^2}
-\frac{1}{(\Delta\tau - i\epsilon)^2 - (2 d_i - kL)^2}\Biggr],
\nn\\
G^{+}[x_i(\tau),x_j(\tau')]\bigr|_{i\neq j} &=& -\frac{1}{4\pi^2}\sum_{k = -\infty}^{\infty}\Biggl[\frac{1}{(\Delta\tau - i\epsilon)^2 - (d_{-} \pm kL)^2}
-\frac{1}{(\Delta\tau - i\epsilon)^2 - (d_{+} - kL)^2}\Biggr],
\label{g2ij}
\nn\\
\eea
We can insert the above results into Eq. (\ref{Fij}) to evaluate the transition rates of the entangled atomic system in the presence of two mirrors.  Hence, the general transition rates in this set up is given by
\bea
{\cal R}_{|\omega'\rangle \to |\omega\rangle} =
- \frac{g^2}{2\pi}\theta(-\Delta\omega)\sum_{k=-\infty}^{\infty}
&\Biggl\{&|m_{\omega \omega'}^{(1)}|^2\left[ \frac{\sin\left(kL \Delta\omega\right)}{kL} -
\frac{\sin\left((2d_1-kL) \Delta\omega\right)}{2d_1-kL}\right]\nn\\
&&\!\!\!\!\!+|m_{\omega \omega'}^{(2)}|^2\left[ \frac{\sin\left(kL \Delta\omega\right)}{kL} -
\frac{\sin\left((2d_2-kL) \Delta\omega\right)}{2d_2-kL}\right]\nn\\
&&\!\!\!\!\!\!\!\!\!\!\!\!\!\!\!\!\!\!\!\!\!\!\!\!\!\!\!\!\!\!\!\!\!\!\!\!\!\!\!\!\!\!\!\!\!\!\!\!\!\!\!\!\!\!\!\!\!\!\!\!\!\!\!\!\!\!\!\!\!\!\!\!\!\!\!\!
+\left[m_{\omega \omega'}^{(1)}m_{\omega \omega'}^{(2)*} +m_{\omega \omega'}^{(2)}m_{\omega \omega'}^{(1)*}\right]
\left[\frac{\sin\left((d_{-}-kL)\Delta\omega\right)}{d_{-}-kL} - \frac{\sin\left((d_{+}-kL)\Delta\omega\right)}{d_{+}-kL}\right]\Biggr\}.
\label{gamma-plane2}
\eea
Each term of the Wightman function in Eq. (\ref{g2ij}) can be written in the frequency domain as
%
\bea
{\cal S}(z,\Delta\omega,L) &=& -\frac{1}{4\pi^2}\sum_{k = -\infty}^{\infty}\int_{-\infty}^{\infty}\,d(\Delta\tau)
\frac{e^{-i \Delta\omega\Delta\tau}}{(\Delta\tau - i\epsilon)^2 - (z - kL)^2},\nn\\
&=& -\frac{\theta[-\Delta\omega]}{2\pi}\sum_{k = -\infty}^{\infty}
\frac{\sin\left[(z - kL)\Delta\omega\right]}{(z - kL)},
\label{integral}
\eea
where we find the contributions to the transition rates replacing in the above expressions the values $z=\{0, 2d_1, 2d_2, d_{-}, d_{+}\}$.
As we are interested in the decay channels $\Delta\omega<0$, thus expanding the summation in Eq. (\ref{integral}) we obtain 
%
\bea
{\cal S}(z,\Delta\omega,L) &= &-\frac{\sin(z\Delta\omega)}{2\pi z}-
\frac{z\sin(z\Delta\omega)}{\pi}\sum_{k = 1}^{\infty}\frac{\cos(kL\Delta\omega)}{z^2-(kL)^2}
+\frac{L\cos(z\Delta\omega)}{\pi}\sum_{k = 1}^{\infty}\frac{k\sin(kL\Delta\omega)}{z^2-(kL)^2}.\nn\\
\label{sumf}
\eea
The series above are further simplified using the relations \cite{prud}
\bea
&&\sum_{k = 1}^{\infty}\frac{\sin(k x)}{k}=\frac{\pi-x}{2}, \,\,\,\,\,\,\,\,\,\,0<x<2\pi,\,\,\\
&&\sum_{k = 1}^{\infty}\frac{\cos(k x)}{k^2- \alpha^2}
=\frac{1}{2\alpha^2}-\frac{\pi}{2}\frac{\cos\left\{\alpha[(2m+1)\pi - x]\right\} }{\alpha\sin(\alpha\pi)}, \,\, 2m\pi\leq x\leq (2m+2)\pi,\,\,\alpha\notin\mathbb{Z},\,\,\\
&&\sum_{k = 1}^{\infty}\frac{k\sin(k x)}{k^2- \alpha^2}
=\frac{\pi\sin\left\{\alpha[(2m+1)\pi - x]\right\} }{2\sin(\alpha\pi)},\,\,\,\,\,\, 2m\pi < x <  (2m+2)\pi,\,\,\, \alpha\notin\mathbb{Z},
\label{sumf2}
\eea
where $m$ is a positive integer. We can prove then, that for all the values of the parameters, the general series reads
\begin{equation}
{\cal S}(z,\Delta\omega,L) =
\begin{cases}
      \hspace{1.6cm} -\Delta\omega/2\pi,         & \frac{z}{L}\in\mathbb{Z},       \quad \hspace{0.8cm}\frac{\Delta\omega L}{\pi}\in\mathbb{Z},\\
      \hspace{1.6cm} -1/2L,                        &  \frac{z}{L}\in\mathbb{Z},      \quad  \hspace{0.8cm}\frac{\Delta\omega L}{\pi}\notin\mathbb{Z},\\
      -\frac{z}{2L^2}\sin(z\Delta\omega)\cot(z\pi/L),   & \frac{z}{L}\notin\mathbb{Z},  \quad  \hspace{0.8cm}\frac{\Delta\omega L}{2\pi}\in\mathbb{Z},\\
      \hspace{0.8cm} -\frac{1}{2L}\frac{\sin[z(2m+1)\pi/L]}{\sin[z\pi/L]}, & \frac{z}{L}\notin\mathbb{Z},  \quad   m<\frac{\Delta\omega L}{2\pi}<m+1.
\end{cases}
  \label{S}
\end{equation}
\begin{figure}
\includegraphics[width=1.0\linewidth]{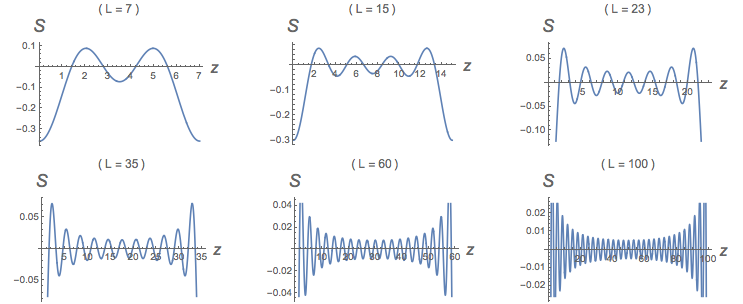}
     \caption{Behavior of the function ${\cal S}(z,\Delta\omega,L)$ defined by Eq. (\ref{S}), for different sizes of the cavity as a function of parameter $z$. We have chosen the energy gap $\Delta\omega=E_{ag}=2.0$ measured in units of $2\pi \lambda_{ga}^{-1}$.} 	
    \label{functionS}
\end{figure}
\begin{figure}
\begin{center}
\includegraphics[width=0.46\linewidth]{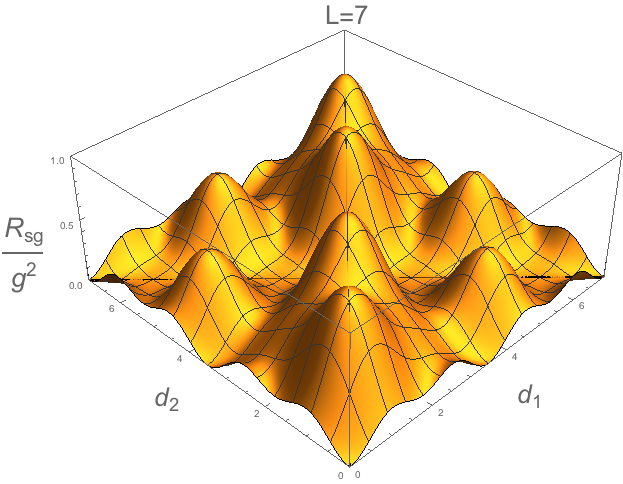}
\includegraphics[width=0.46\linewidth]{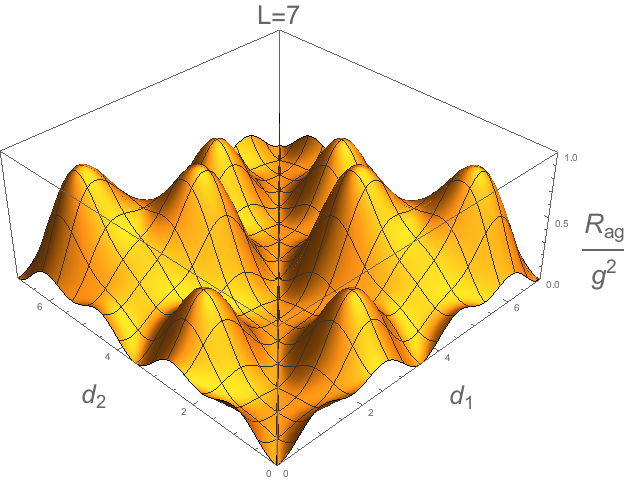}\\
\includegraphics[width=0.49\linewidth]{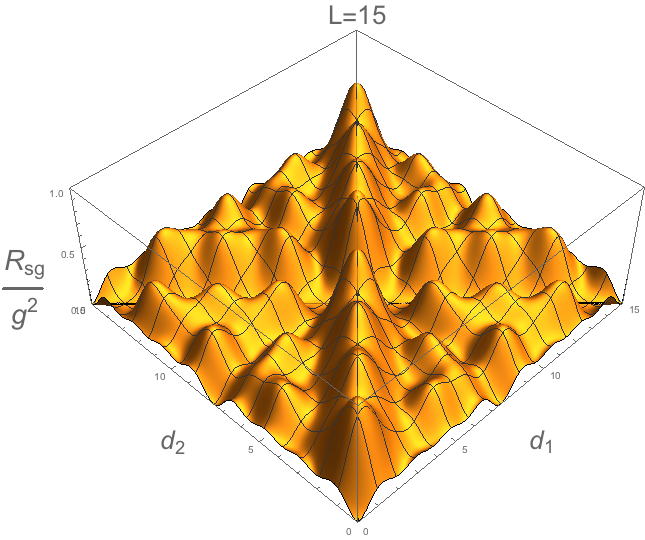}
\includegraphics[width=0.49\linewidth]{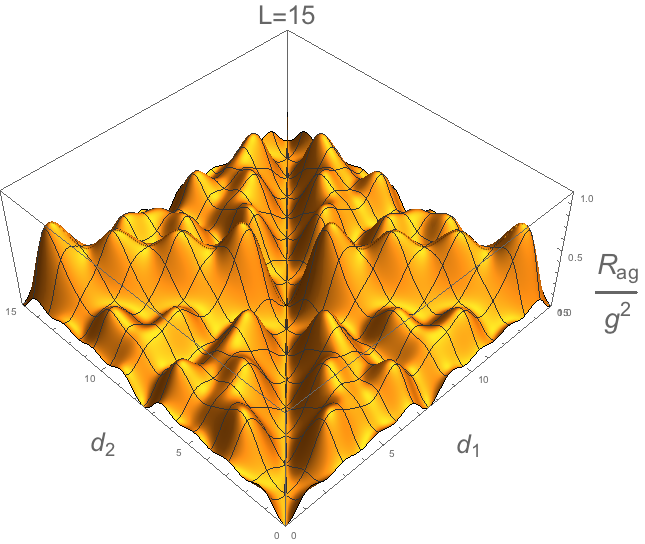}\\
\includegraphics[width=0.49\linewidth]{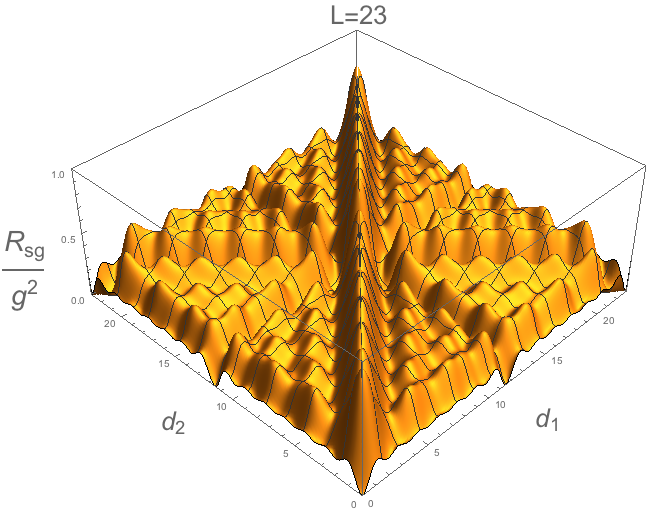}
\includegraphics[width=0.49\linewidth]{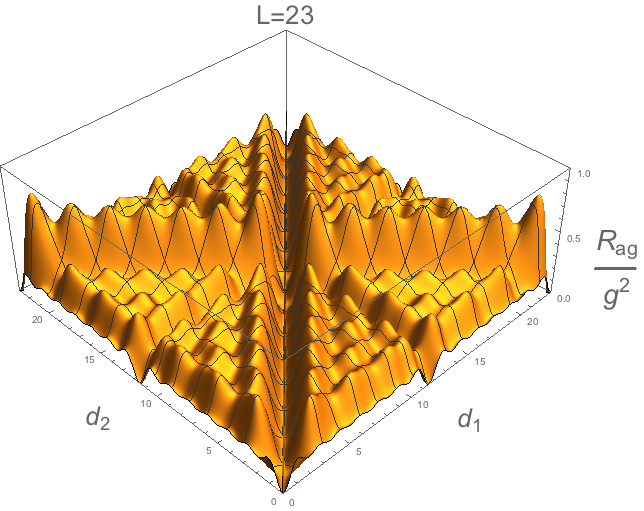}	
\end{center}
     \caption{Spontaneous transition rates ${\cal R}_{sg}$ (\textit{left}) and ${\cal R}_{ag}$ (\textit{right}) as functions of the atom's positions $d_1$ and $d_2$ from the $x_3 = 0$-plane inside the cavity. Upper plots are for a distance between mirrors equals to $L=7$, middle plots for $L=15$, and bottom plots for $L=23$. There is a complete inhibition of the transition rate for the symmetric state if both atoms are located in a symmetric way with respect to the center of the cavity and close to the mirrors. For the antisymmetric state the transition rate vanishes if the atoms are close enough. 
     Again the energies and distances are given in terms of the natural units associated with each transition as in the case of one mirror.} 	
    \label{L8}
\end{figure}
%
This function is shown in the Fig.~(\ref{functionS}) for different sizes of the cavity as a function of parameter $z$. This is a continuous and symmetric function with respect to the midpoint between the boundaries. We see a sinusoidal behavior depending on the size of the cavity being small at its center. This feature characterizes the profile of the spontaneous transition rate for collective states as function of the positions of the atoms inside the cavity. With the result obtained in Eq. (\ref{S}), we can find the transition rate for any specific state. Let us again consider the two transitions decay from the maximally entangled states as before.

\textit{Symmetric state transition}: The spontaneous decay rate of the two-atoms system from the symmetric entangled state $|s\rangle$ to the ground state $|g\rangle$ in the presence of two infinite plane mirrors reads
\bea
{\cal R}_{|s\rangle \to |g\rangle}=g^2\biggl\{{\cal S}(0,|E_{gs}|,L)&-&\frac{1}{2}{\cal S}(2d_1,|E_{gs}|,L)
-\frac{1}{2}{\cal S}(2d_2,|E_{gs}|,L) \nn\\
&+&{\cal S}(d_-,|E_{gs}|,L) - {\cal S}(d_+,|E_{gs}|,L)\biggr\}.
\label{R2sg}
\eea
The results of this transition rate for different values of the distance between the plates are shown on the left sides of
Fig. (\ref{L8}). In all of these plots we note that for symmetrical positions of the atoms with respect to the center of the line perpendicular to the mirrors, $d_1+d_2=L$, the decay rate is zero. This means that the symmetric state remains stationary and unperturbed by the vacuum fluctuations if the atoms are located symmetrically inside the cavity formed by the mirrors.

\textit{Antisymmetric state transition}: Let us investigate the transition from the antisymmetric entangled state $|a\rangle$ to the ground state $|g\rangle$. Considering the corresponding matrix elements of this transition, we obtain the decay rate inside the two-mirrors cavity given by
\bea
{\cal R}_{|a\rangle \to |g\rangle}=g^2\biggl\{{\cal S}(0,|E_{ga}|,L)&-&\frac{1}{2}{\cal S}(2d_1,|E_{ga}|,L)
-\frac{1}{2}{\cal S}(2d_2,|E_{ga}|,L)\nn\\
&-&{\cal S}(d_-,|E_{ga}|,L)+{\cal S}(d_+,|E_{ga}|,L)\biggr\}.
\label{R2ag}
\eea
The results of this transition rate for different values of $L$ are shown on the right side of Fig. (\ref{L8}). 
We can understand these last results by noting that the Wightman function for the case of two mirrors, Eq. (\ref{wighp2}), has the property
\begin{eqnarray}
G^+[\Delta\tau,\Delta{\bf x}_{\perp},x_3;x'_3]&=&- G^+[\Delta\tau,\Delta{\bf x}_{\perp},x_3;L- x'_3],\nonumber\\
&=&-G^+[\Delta\tau,\Delta{\bf x}_{\perp},L-x_3;x'_3].
\label{eq:property}
\end{eqnarray}
This property shows the reflection symmetry of the system in the $x_3$-direction and characterizes the relation between the transition rate profiles for entangled states. Using the Eq. (\ref{pro2}), Eq. (\ref{Fij}) and Eq. (\ref{mel}), we can write the transition rates for both entangled states as
%
\begin{eqnarray}
\!\!\!\!\!\!\!\!\!\!\!\!\!\!\! {\cal R}_{ag}(|E_{ga}|,d_1,d_2) &=& \frac{g^2}{2}\bigg[ {\cal F}_{11}(|E_{ga}|,d_1) + {\cal F}_{22}(|E_{ga}|,d_2) \nonumber\\
&-& {\cal F}_{21}(|E_{ga}|,d_2,d_1) - {\cal F}_{12}(|E_{ga}|,d_1,d_2)\bigg],
\end{eqnarray}
\begin{eqnarray}
{\cal R}_{sg}(|E_{gs}|,d_1,d_2) &=& \frac{g^2}{2}\bigg[ {\cal F}_{11}(|E_{gs}|,d_1) + {\cal F}_{22}(|E_{gs}|,d_2) \nonumber\\
  &&\!\!\!\!\!\!\!\!\!\!\!\! - {\cal F}_{21}(|E_{gs}|,L-d_2,d_1) - {\cal F}_{12}(|E_{gs}|,d_1,L-d_2)\bigg] ,
\end{eqnarray}
where in the last equation we used the Eq. (\ref{eq:property}).
These results unveil us two features. First, omitting the dependence on $|E_{gs}|$ in ${\cal F}_{ij}$, if both atoms are very close to each other ($d_1\approx d_2=d$) in the antisymmetric state ${\cal F}_{11}\approx {\cal F}_{22}$ and ${\cal F}_{12}\approx {\cal F}_{21}\approx {\cal F}_{11}$, thus ${\cal R}_{ag}(d,d)\rightarrow 0$ (see Fig. (\ref{L8}, right)). For the symmetric case something similar takes place if the atoms are located at symmetric positions with respect to the center of the cavity, i.e., $d_1+d_2 = L$. In this case, since ${\cal F}_{11}(d_1)={\cal F}_{22}(d_1)$, then ${\cal F}_{22}(L-d_1)=-{\cal F}_{22}(d_1)=-{\cal F}_{11}(d_1)$. 
Also we have that ${\cal F}_{12}(d_1,L-d_1)=-{\cal F}_{12}(d_1,d_1)=-{\cal F}_{11}(d_1)$ and for a similar reason ${\cal F}_{21}(d_2,L-d_2)=-{\cal F}_{22}(d_2)={\cal F}_{22}(L-d_2)={\cal F}_{22}(d_1)$ which implies that ${\cal R}_{sg}(d,L-d)\rightarrow 0$, (see Fig. (\ref{L8}, left)). At these configurations, it is possible to verify that the antisymmetric state $|a\rangle$ (for atoms sufficiently close to each other) and the symmetric state $|s\rangle$ (for atoms at symmetric positions with respect to the center of the cavity) are eigenstates of the total Hamiltonian (including the interaction with the field), and therefore the presence of the interaction does not change the stationary feature of these states. 

On the other hand, 
to show how the transition rates are related, we focus on the top-right plot of Fig. (\ref{L8}) for the antisymmetric state. We can dislocate the origin of the plot to the end of the $d_2$-axis, then we reverse the direction of this axis and turn up to down the plot to match the orientations of the axis with those of the top-left plot of the Fig. (\ref{L8}). Indeed, we will obtain exactly the transition rate for the symmetric state. It shows precisely the property of the correlation function Eq. (\ref{eq:property}). Thus, the reflection symmetry of the system relates the transition rates for the maximally entangled states.
%
%
These considerations imply that we have a complete inhibition of the spontaneous transition rate for two distinct situations. We remark that from Fig. (\ref{L8}), we see that even if one of the atoms is placed on one of the mirrors (for instance $d_1=0$) where the field vanishes, there exists a probability for the transition to take place. Moreover, this probability can be greater than the probability evaluated for atoms located inside the cavity. 

\newpage
\section{Conclusions and perspectives}
\quad

By using a first-order approximation in time-dependent perturbation theory, we investigated radiative processes of two-level atoms in an entangled state interacting with a massless scalar field. We studied the transition probability per unit proper time for inertial atoms in empty space and also in the presence of boundaries. In the former, we see that the spontaneous decay rates can be enhanced or inhibited depending only on the specific entangled state and on the separation between the atoms. For the symmetric state, if both atoms are separated by distances smaller than the wavelength associated with the transition energy gap, the spontaneous transition rate will be enhanced with respect to the case where the entangled atoms are separated by large distances. For the antisymmetric state for such distances the spontaneous decay rate presents a total inhibition. It implies that for the former case the quantum cross-correlations generate a constructive interference, whereas in the latter case the interference is destructive. For large distances compared with the resonant wavelength of the transitions, the spontaneous decay rates present a decreasing sinusoidal behavior. 

For the case of a single mirror, the decay rates of the atoms from the maximally entangled states are slightly modified and described by the expressions Eq.~(\ref{gamma-plane-sg}) and Eq.~(\ref{gamma-plane-ag}). Therefore, the mirror can enhance or decrease the spontaneous transition rates depending on the entangled state considered and the relative positions of the atoms with respect to the mirror. 
For the case of two parallel mirrors we obtained that the transition rates of the two-atom system are summarized in Eq.~(\ref{R2sg}) and Eq.~(\ref{R2ag}). These decay rates from the entangled states of the system are presented in Fig.~(\ref{L8}). There we see that the transitions rates from the maximally entangled states form patterns of interference inside the cavity. Therefore, depending of the size of the cavity and the relative positions of both atoms inside the cavity there exist a series of maxima and minima for the decay rates.

On the other hand, we are aware that recent results show that in the $(1+1)$-dimensional case long-time behavior of the dynamics of atoms will be largely altered by the echoes of the retarded quantum field emitted by the atoms~\cite{ref1,ref2,ref3}. The effects on higher dimensional space-time deserve further investigations. However, in the situation studied in the present paper we adopt a conservative approach which ignores the back-reactions. This amounts to consider our atoms as point-like objects.

A natural generalization of this paper is the investigation of atoms interacting with the electromagnetic field using the Heisenberg picture~\cite{dali1,dali2}. This approach allows an easy comparison of quantum mechanical and classical concepts. In these studies, the notion of vacuum fluctuations was connected with the free solutions of the Heisenberg equations for the quantum field. Radiation reaction was incorporated via the source field, which is the part of the field caused by the presence of the atom itself. An interesting application of such formalism can be found in Ref.~\cite{aud}. In this context, one can analyze the contributions of vacuum fluctuations and radiation reaction to the disentanglement of the same atomic system considered here~\cite{ng1}. It is interesting to ask how this formalism can be employed in the situation of atoms confined in a cavity.

The notion of locality by Einstein implies that no information can travel faster that the speed of light. This idea appears in the quantum field theory in the construction of the $S$ matrix \cite{stueckelberg}. Let us we consider processes in which transitions occur in different space-time regions $\Omega_x$ and $\Omega_y$ separated by a distance $R$.  If the energy of the system defined inside $\Omega_x$ decreases by $\omega_0$ at time $t = 0$ accompanied by the increase of the energy of the system defined inside $\Omega_y$ by the same amount $\omega_0$ at time $t$, then we must have that $t > R/c$. This problem was discussed in the $1930's$ by Fermi. An interesting possibility is to discuss the Fermi problem with a different experimental set up  \cite{fermi, Hegerfeldt, Buchholz, Borrelli, Jonsson}. The idea is to investigate causality problems in the system which consists of a free atom prepared in a ground state and two atoms prepared in an entangled state. The two entangled atoms and the free atom are localized in disjoint regions separated by a distance $d$.
These subjects are under investigation by the authors.

\section*{Acknowlegements}

We would like to thank Jorge Stephany Ruiz and Tobias Micklitz for useful discussions, and the referees for their suggestions to improve our work. This paper was supported by Brazilian agencies: Conselho Nacional de Desenvolvimento Cientifico e Tecnol{\'o}gico do Brasil (CNPq) and Coordena\c{c}\~ao de Aperfei\c{c}oamento de Pessoal de N\'ivel Superior (CAPES).


\end{document}